\documentclass[a4paper,11pt]{article}
\usepackage{pos}

\usepackage{amsmath}
\usepackage{amssymb}
\usepackage{subcaption}
\usepackage{float}
\usepackage{braket}
\usepackage{wrapfig}
\usepackage{bbold}
\usepackage{hyperref}
\usepackage{bm}
\usepackage{lineno}
\usepackage{multirow}

\usepackage{graphicx} 
\usepackage{booktabs} 
\usepackage{subcaption}

\title{Towards $K\pi$ scattering with domain-wall fermions at the physical point using distillation}

\author*[a]{Nelson Pitanga Lachini}
\author[a,b]{Peter Boyle}
\author[a]{Felix Erben}
\author[a]{Michael Marshall}
\author[a]{Antonin Portelli}

\affiliation[a]{School of Physics and Astronomy, University of Edinburgh, Edinburgh EH9 3JZ, UK,}
\affiliation[b]{Brookhaven National Laboratory, Upton, NY 11973, USA}

\emailAdd{nelson.lachini@ed.ac.uk}

\abstract{Resonances play an important role in Standard Model phenomenology. In particular, hadronic resonances feature in $B$ and $D$ decays, which can be central for New Physics searches. Lattice QCD simulations combined with the finite-volume method can nowadays be used to reliably study strongly coupled scattering processes such as $K\pi$ and thus the hadronic resonance $K^*$. In this work, we approach $K\pi$ scattering on a domain-wall $N_f = 2+1$ RBC-UKQCD
ensemble at a physical pion mass. We use the distillation method within Grid and Hadrons software to compute sets of operator basis. That allows solving an eigenvalue problem to extract the low-energy finite-volume spectra, which are then translated into scattering information. We update the state of the calculation by reviewing the smearing process, outlining the variational analysis and concluding by showing preliminary data.}

\FullConference{%
  The 39th International Symposium on Lattice Field Theory (Lattice2022),\\
  8-13 August, 2022 \\
  Bonn, Germany 
}

\begin{document}
\maketitle

\section{Introduction}

Lattice QCD is formulated on Euclidean space-time and simulations always take place on finite volumes, which means that there is no way of obtaining infinite-volume scattering information directly from the lattice. We only have access to the Euclidean spectrum of the theory. Nevertheless, an indirect connection between the spectrum of a finite-volume Euclidean theory and the scattering amplitude of the correspondent infinite-volume Minkowski theory was established already in the 1980s \cite{Luscher1986,Luscher1986a} and was further developed over the years \cite{Rummukainen1995,Kim2005a,Hansen2012}. Moreover, lattice QCD is able to address hadronic resonances by studying the scattering phase shift obtained from such finite-volume analysis \cite{Luscher1991a}.

The number of lattice studies of hadronic resonances has increased over the last years but still is in a development stage \cite{Briceno2018}. As we approach physical pion masses at fixed $m_\pi L \sim 4$, the number of energy levels in the regime of elastic scattering decreases. This further motivates the inclusion of moving frames on the calculation in order to satisfactorily constrain scattering amplitude parametrisations. In this work we will be interested in studying $I=1/2, $ $P$-wave $K \pi$ scattering and thus the $K^*(892)$ resonance at a physical pion mass. This calculation was addressed previously at higher pion masses by other collaborations \cite{Rendon2020,Brett2018,Prelovsek2013a,Bali2016}.

Another clear difficulty of scattering studies at physical pion mass is its computational cost. As we decrease $m_{\pi}$, the physical spatial extension $L$ must be increased to keep exponentially suppressed finite-volume effects under control. For scattering studies, it is crucial to have access to a reasonable operator basis in order to use variational analysis methods, and for this we use distillation to smear quark fields and compute lattice correlators \cite{Peardon2009,Morningstar2011}. 

In a previous study \cite{Lachini2022}, using an RBC-UKQCD $N_f=2+1$ domain-wall fermion lattice with $m_{\pi}\approx 139$ MeV and $m_K\approx 499$ MeV and a $48^3 \times 96$ volume \cite{Blum2016}, we tuned the distillation setup based on the smearing profile and signal of simple correlators. There, we concluded that using exact distillation with $64$ eigenvectors of the $3D$-covariant Laplacian was the best compromise for carrying out this $K\pi$ scattering calculation.

We use Grid as the data parallel C++ library for the lattice computations and Hadrons as the workflow management system for the measurements \cite{Boyle2015,Hadrons2022}, open-source and free software. The efficient computation of meson fields at manageable storage cost demanded the writing of dedicated distillation code within Grid and Hadrons, which is being documented as well \cite{DistilDoc2021}. Such distillation code was also showcased in Refs. \cite{Boyle2019,Lachini2022,Fabian2022}.

\section{Distillation}

To compute the correlator data necessary to extract the finite-volume energy spectrum, we use the so-called distillation method \cite{Peardon2009,Morningstar2011}. Distillation involves a combination of link smearing and $3D$-Laplacian (Lap) quark smearing, which are both gauge covariant by construction. Given the gauge-covariant $3D$-Laplacian $\bm \nabla^2$ \cite{Peardon2009} eigenvalues and eigenvectors on a certain time slice, namely
\begin{equation}
- \bm \nabla^2 v_k(t) = \lambda_k (t) v_k(t), \quad k=1, 2, \ldots, \qquad 0 < \lambda_1 < \lambda_2 < \ldots \ .
\end{equation}
the distillation smearing operator is defined as a projector onto the low-mode subspace of $- \bm \nabla^2$, i.e.
\begin{equation}
\label{distillationkernel}
  \mathcal{S} (t)  = \sum_{k=1}^{N_{\mathrm{vec}}} v_k (t)  v_k (t)^{\dagger}.
\end{equation}  
The distillation operator defines the smeared quark field through $\psi(t) \to \mathcal{S} (t) \psi(t)$. The suppression of short-distance modes is desirable as they do not affect low-energy physical signals in hadron correlation functions.

From appropriately defined solve (or sink) $\varphi$ and source $\varrho$ vectors \cite{Lachini2022}, we can write an estimator for the quark propagator as
\begin{equation}
    S_{\mathbf x \mathbf y}(t_f,t_0) = \varphi(\mathbf x,t_f) \varrho(\mathbf z,t_0)^{\dagger}.
\end{equation}
Exact distillation can effectively be described using these distillation objects by using trivial "noise" vectors (all entries equal to one) and full-dilution in all indices. We can then write arbitrary traces of propagators in terms of $\varphi$ and $\varphi$, e.g.
\begin{equation}
    \label{simpletrace}
    tr [\Gamma S(x,y) \Gamma' S'(y,x)] = tr [ \varrho(y)^{\dagger} \Gamma \varphi(y) \ \varrho(x)^{\dagger} \Gamma' \varphi'(x)] = tr [ M_{\Gamma}(\varrho,\varphi;y) \ M_{\Gamma'}(\varrho,\varphi';x) ],
\end{equation}
where $M_{\Gamma}(\varrho\varphi;x) = \varrho(x)^{\dagger} \Gamma \varphi(x)$ are the building blocks of correlation functions, called meson fields.

Following the distillation workflow as outlined in Ref. \cite{Lachini2022}, we computed strange and light inversions on every time slice. Then we generate meson fields of the kind $M(\varrho\varphi;x)$. Based on the study done in Ref. \cite{Lachini2022}, we carry out all measurements using $N_{\mathrm{vec}}=64$.

\section{Variational Analysis}

To perform a lattice scattering calculation based on the finite-volume formalism, we need to obtain towers of low-lying energies in several moving frames. Defining a variational problem and solving a generalized eigenvalue problem (GEVP) applied to a correlator matrix is a possible way to proceed \cite{Briceno2018}. Multi-hadron correlators are suitable for being computed within distillation, as we can combine the same Laplacian-projected propagators with different interpolators and generate correlators with potentially different overlaps to the various states of interest.

We use an operator basis containing bilinear ($\bar \psi \Gamma \psi'$) and two-hadron operators. In particular, we use $O_{V} = \bar s \gamma^i u$ as the single vector interpolator. For the pseudoscalar states, we use the conventional $\pi^+, \pi^0, K^+, K^0$ interpolators with $\Gamma=\gamma^5$, which combined and projected to isospin $I=1/2, I_3=1/2$, yield
\begin{equation}
    O_{K \pi}(\mathbf p_1, \mathbf p_2)  = \sqrt{\frac{1}{3}} O_{K^+}(\mathbf p_1) O_{\pi^0}(\mathbf p_2) + \sqrt{\frac{2}{3}} O_{K^0}(\mathbf p_1) O_{\pi^+}(\mathbf p_2).
\end{equation}
For a moving frame with total spatial momentum $\mathbf P$, the individual momenta obey $\mathbf P = \mathbf p_1 + \mathbf p_2$. All operators obeying $|\mathbf P|^2 \le 4$ and $|\mathbf p_1|^2, |\mathbf p_2|^2 \le 4$ in units of $2\pi/L$ were included.

These operators must transform according to the irreducible representations $\Lambda$ of the appropriate little group $G$ on the lattice. This is done through the projection formula in momentum space \cite{Gockeler2012a}
\begin{equation}
    P^{\Lambda} \mathcal{O}(\mathbf p_1,\mathbf p_2,...) = \frac{d_{\Lambda}}{n_G} \sum_{i=1}^{n_G} \chi_{(i)}^{\Lambda*} \ \hat S^{(i)} \mathcal{O}(S^{(i)} \mathbf p_1, S^{(i)} \mathbf p_2,...) \hat S^{(i) \dagger} \equiv O^{\Lambda}(\mathbf p_1,\mathbf p_2,...),
\end{equation}
where $S$ are the elements of the little group $G$ of order $n_G$. Such formula uses the character $\chi^{\Lambda}$ of irrep $\Lambda$ with dimension $d_{\Lambda}$.

In this work, we use only the irreps $\Lambda$ with a leading occurrence in continuum $P$-wave ($J=1$ irreps) and where odd and even partial waves do not mix \cite{Leskovec2012a}, i.e. $\{\mathbf P, \Lambda\} = \{[000],T_{1u}\}$, $\{[001],E\}$, $\{[110],B_1\}$, $\{[110],B_2\}$,  $\{[111],E\}$ and $\{[002],E\}$.

\begin{table}[H]
    \centering
    \begin{tabular}{|c|c|c|c|}
        \hline
        \bm{$G$}              & \bm{$P$}           & \bm{$\Lambda$} & \bm{$J$} \\ \hline
        $O_h$                     & [000]                  & $T_{1u}$           & $1,3,\ldots$ \\ \hline
        $C_{4v}$                  & [001],[002]            & $E$                & $1,2,\ldots$ \\ \hline
        \multirow{2}{*}{$C_{2v}$} & \multirow{2}{*}{[110]} & $B_1$              & $1,2,\ldots$ \\ \cline{3-4} 
                                  &                        & $B_2$              & $1,2,\ldots$ \\ \hline
        $C_{3v}$                  & [111]                  & $E$                & $1,2,\ldots$ \\ \hline
    \end{tabular}
    \caption{Occurrence of continuum total angular momentum $J$ in the $O_h$ subgroup irreps used here \cite{Gockeler2012a}.}
\end{table}

\begin{figure}[H]
	\centering
	\includegraphics[width=0.99\linewidth]{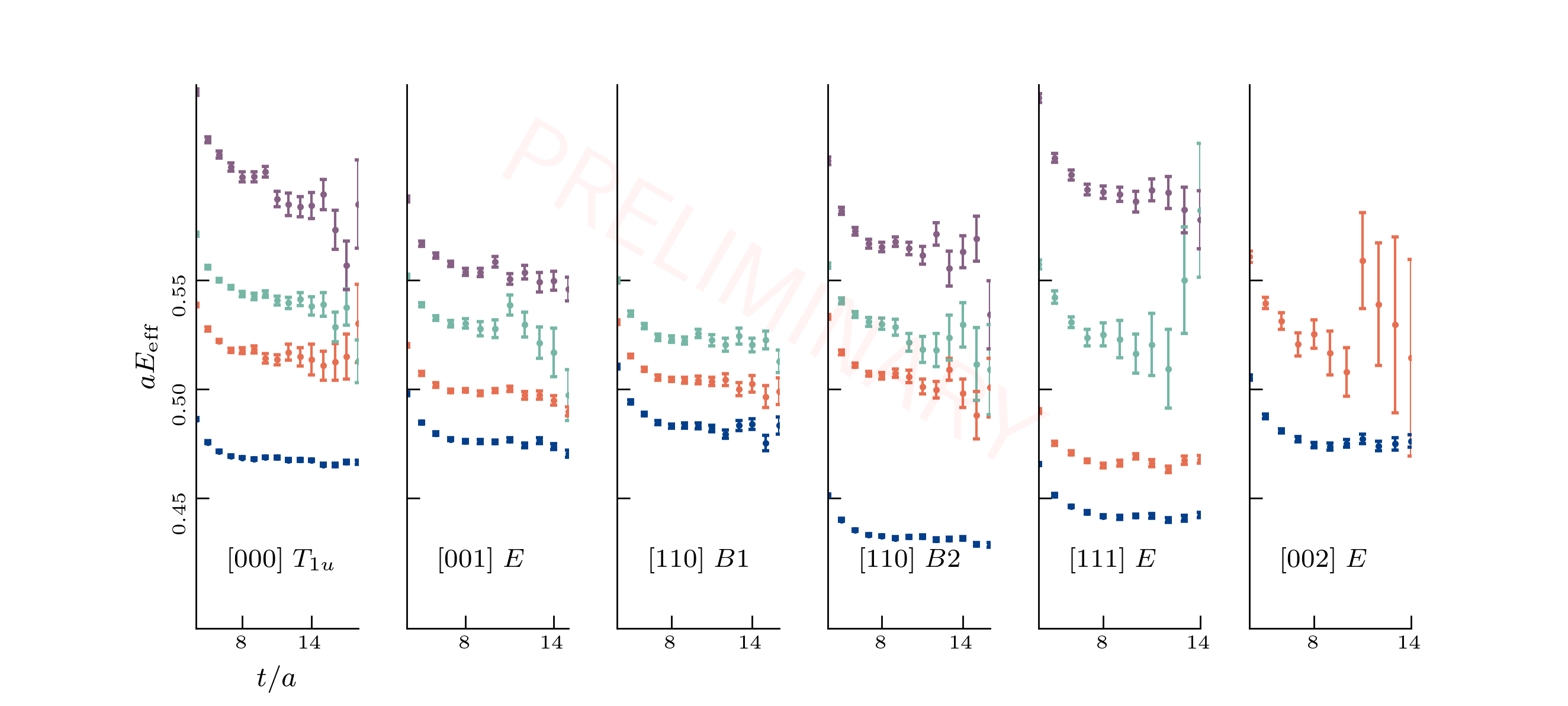}
	\caption{Effective energy plot of GEVP correlators in all irreps considered. The energy axis is kept fixed along all frames, but the time axis is shifted. We removed the highest level of each GEVP, as this is the one receiving the uncontrolled excited state contributions \cite{Blossier2009}. }
	\label{fig:gevp}
\end{figure}

The GEVP was solved using a fixed-$t_0$ method defined by 
\begin{equation}
    C(t) u^{(n)}(t,t_0) = \lambda^{(n)}(t,t_0) C(t_0) u^{(n)}(t,t_0),
\end{equation}
where, after some testing, we chose to take $t_0=3$. Solving this equation and using effective energies of the form 
\begin{equation}
    a E^{(n)}_{\mathrm{eff}}(t) = \log \frac{\lambda^{(n)}(t,t_0)}{\lambda^{(n)}(t+1,t_0)}
\end{equation}
yield the levels in Fig. \ref{fig:gevp}. 

Note that, apart from the lowest levels in $\{[110],B_2\}$ and $\{[111],E\}$, all energies are above the $K\pi\pi$ threshold, which means that in principle they would not be treatable by the $2$-particle Lüscher's analysis. However, we stress that the experimental $K^*(892)\to K\pi\pi$ decay fraction is $\mathcal O(10^{-4})$ \cite{PDG2022}, meaning that the $3$-particle corrections will most probably be negligible. This issue will be addressed in more detail in a later.

\begin{figure}[H]
	\centering
	\includegraphics[width=.99\linewidth]{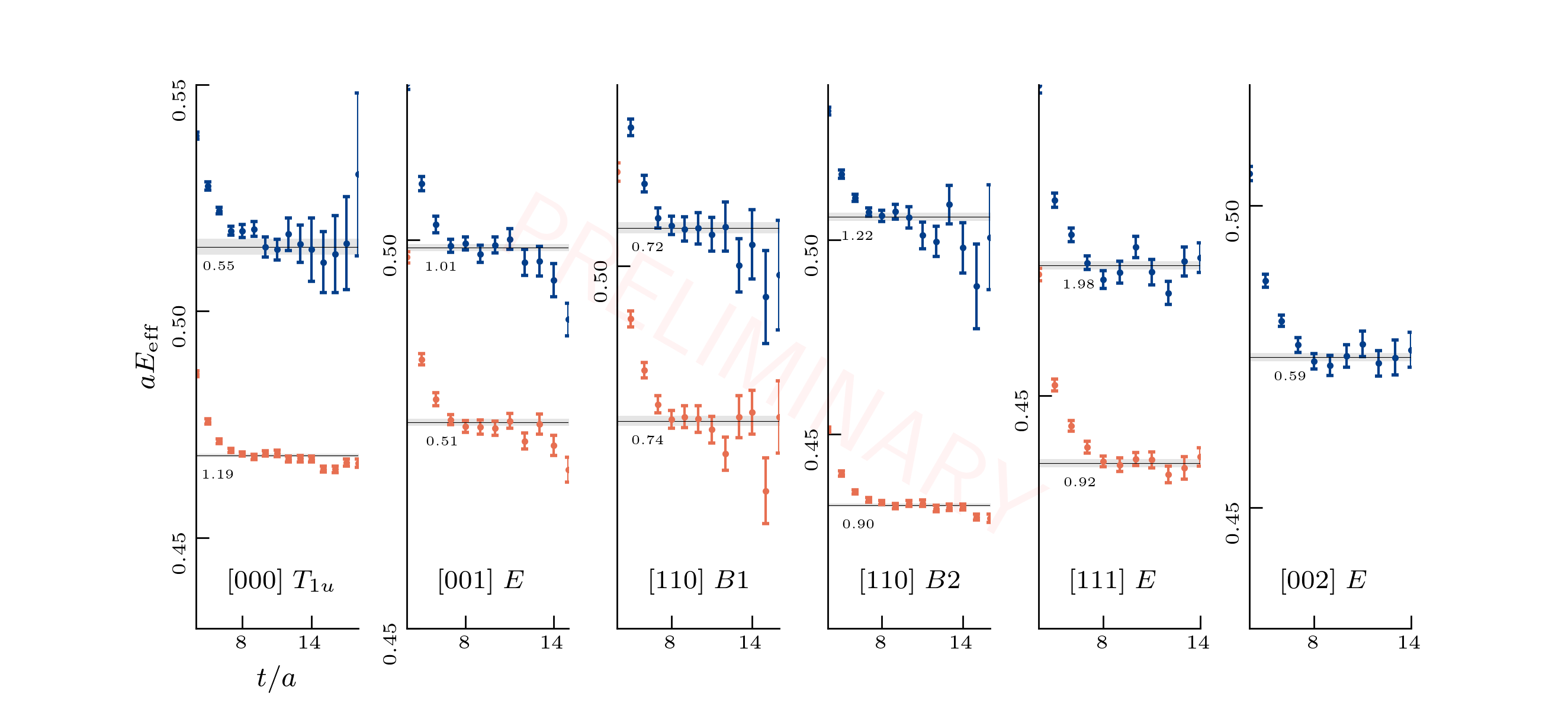}
	\caption{Best AIC fit results (grey band) overlaid with respective effective energies on the lowest GEVP levels. The $\chi^2/ n_{dof}$ is shown below each fit.}
	\label{fig:gevpfit}
\end{figure}

To extract the energy levels from the GEVP data, we do correlated fits of each eigenvalue $\lambda^{(n)}(t,t_0)$ to a single exponential. The choice of fit range is done by a scan over the time extension with an acceptable signal-to-noise ratio. Then, using an AIC weighting of the correlated fits \cite{Akaike1974,Andrew2022}, we choose fit ranges yielding the results illustrated in Fig. \ref{fig:gevpfit}, where we picked only levels with reasonable quality of fit. At this level of statistics, we also do a thorough visual inspection of the effective energies plateaus against the selected fit ranges.

\section{Finite-volume Analysis}

In our case, the $2$-particle quantisation condition for pure $P$-wave scattering for each moving frame and irrep can be written in the form \cite{Leskovec2012a}
\begin{equation}
    \label{eq:qc}
    \tan \delta_1 (p^*) = \phi^{\mathbf P, \Lambda}(q),
\end{equation}
where $p^*= |\mathbf p^*|$ refers to the center-of-mass momentum of $K$ and $\pi$, and $q = p^* L / (2\pi)$. The finite-volume function $\phi^{\mathbf P, \Lambda}$ depends on the generalized zeta function, which can be evaluated numerically.

\begin{figure}[H]
	\centering
	\includegraphics[width=.99\linewidth]{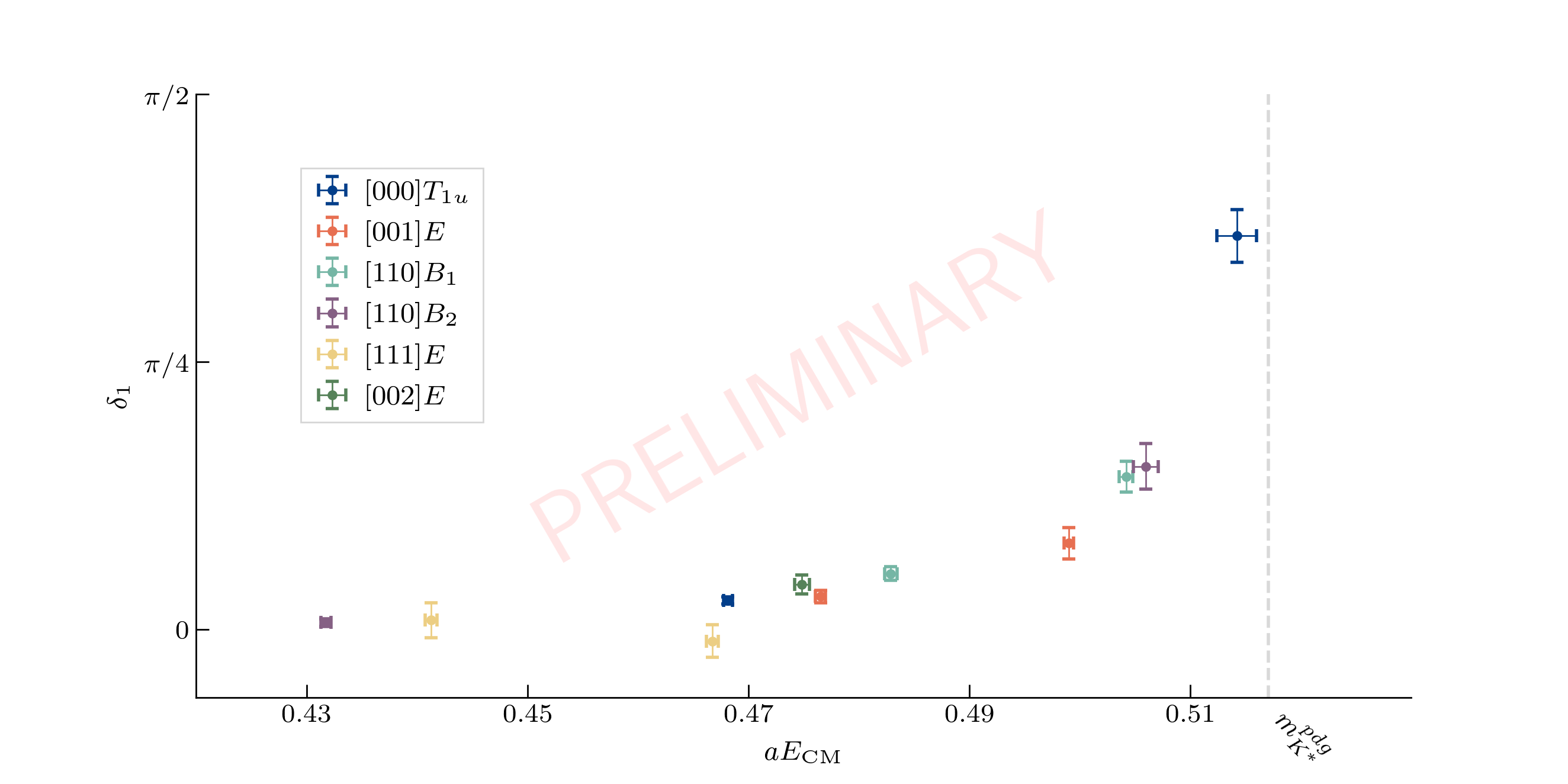}
	\caption{Phase shift plotted against CoM energy obtained from equation \eqref{eq:qc}, with irreps and moving frames distinguished. The dashed vertical line represents the experimental value for the $K^*(892)$ mass in lattice units \cite{PDG2022}.}
	\label{fig:phaseshift}
\end{figure}

By feeding the results illustrated in Fig. \ref{fig:gevpfit} into equation \eqref{eq:qc}, we obtain the preliminary $P$-wave phase shift plot of $I=1/2$ $K\pi$ scattering at a physical pion mass in Fig. \ref{fig:phaseshift}.

\section{Conclusions and Outlook}

Using exact distillation at $N_{\mathrm{vec}}=64$, we have computed multi-hadron correlators suitable for a $K \pi$ scattering study at physical pion mass. We have performed a GEVP to all pure $P$-wave irreps on moving frames with total momenta up to $|\mathbf P|^2 = 4 (2\pi/L)^2$ and performed correlated fits of the eigenvalues to a single exponential. The energy levels obtained were fed to the $2$-particle quantisation, yielding the preliminary phase shift results for $I=1/2,$ $K\pi$ scattering.

As an immediate next step, we will increase statistics to cover the whole physical point ensemble used here,  allowing for better control of the fits. This will enable a reliable extraction of $K^*(892)$ resonance parameters by an appropriate fit to the spectrum. In future work, we will also include the irreps mixing $S$ and $P$ waves, where the so-called $\kappa$ resonance is expected to show up.

\vspace{1cm}

\textbf{Acknowledgements} 

The authors thank the members of the RBC and UKQCD Collaborations for the helpful discussions and suggestions.

N.L., F.E. and A.P. also kindly thank Mike Peardon for the invaluable discussions.

This work used the DiRAC Extreme Scaling service at the University of Edinburgh, operated by the Edinburgh Parallel Computing Centre on behalf of the STFC DiRAC HPC Facility (\href{https://dirac.ac.uk/}{\texttt{https://dirac.ac.uk/}}). The equipment was funded by BEIS capital funding via STFC grants ST/R00238X/1 and STFC DiRAC Operations grant ST/R001006/1. DiRAC is part of the National e-Infrastructure.

PB has been supported in part by the U.S. Department of Energy, Office of Science, Office of Nuclear Physics under the Contract No. DE-SC-0012704 (BNL). P.B. has also received support from the Royal Society Wolfson Research Merit award WM/60035.

N.L. \& A.P. received funding from the European Research Council (ERC) under the European Union’s Horizon 2020 research and innovation programme under grant agreement No 813942.

M.M. gratefully acknowledges support from the STFC in the form of a fully funded PhD studentship.

A.P. \& F.E. are supported in part by UK STFC grant ST/P000630/1. A.P. \& F.E. also received funding from the European Research Council (ERC) under the European Union’s Horizon 2020 research and innovation programme under grant agreements No 757646.

\bibliographystyle{JHEP}
\bibliography{library}

\end{document}